\documentclass[10pt]{iopart}
\usepackage{graphicx}
\usepackage{amsfonts}
\usepackage{color}
\newcommand {\be}{\begin{equation}}
\newcommand {\ee}{\end{equation}}

\begin{document}

\paper[]{Hydrodynamics and transport \\in the long-range-interacting $\varphi^4$ chain}
\date{\today}

\author{Stefano Iubini$^{1,2}$, Stefano Lepri $^{1,2}$ and Stefano Ruffo $^{3,1}$\
}
\address{$^{1}$  Consiglio Nazionale
delle Ricerche, Istituto dei Sistemi Complessi, 
via Madonna del Piano 10, I-50019 Sesto Fiorentino, Italy
}
\address{$^{2}$Istituto Nazionale di Fisica Nucleare, Sezione di Firenze,  
via G. Sansone 1 I-50019, Sesto Fiorentino, Italy}

\address{$^{3}$  SISSA and INFN,
Sezione di Trieste,
Via Bonomea 265, I-34136 Trieste,
Italy 
}
\ead{stefano.iubini@isc.cnr.it,stefano.lepri@isc.cnr.it,ruffo@sissa.it\newline 
}

\begin{abstract}
We present a simulation study of the one-dimensional  $\varphi^4$ lattice 
theory  with long-range interactions decaying as an inverse power 
$r^{-(1+\sigma)}$ of 
the intersite distance $r$, $\sigma>0$.  We consider the cases of 
single and double-well
local potentials with both attractive and repulsive couplings. 
The double-well, attractive case displays a phase transition
for $0<\sigma \le 1$
analogous to the Ising model with long-range ferromagnetic interactions.
A dynamical scaling analysis of both energy structure 
factors and excess energy correlations shows that the effective hydrodynamics
is diffusive for $\sigma>1$ and anomalous for $0<\sigma<1$ where 
fluctuations propagate superdiffusively. We argue that this is 
accounted for by a fractional diffusion process and we compare the results 
with an effective model of energy transport based on L\'evy flights. 
Remarkably, this 
result is fairly insensitive on the phase transition. 
Nonequilibrium simulations with an applied thermal gradient 
are in quantitative agreement with
the above scenario.  
\end{abstract}

\section{Introduction}

Nonequilibrium properties of many-body statistical systems on macroscopic 
scales are usually well described by hydrodynamic equations.
This is justified by the fact that macroscopic fluctuations of conserved quantities
(and order parameters close to criticality) 
must evolve slowly with respect to microscopic time-scales  
to reach a steady state. If the relevant correlations have long-time tails,
standard diffusive behavior may break down as it happens generically in 
nonlinear, low-dimensional systems \cite{LLP03,DHARREV,Lepri2016,Benenti2020}.
Anomalous transport in such many-body systems can be effectively described
by a random L\'evy walk \cite{Zaburdaev2015} of the energy carriers, as demonstrated extensively in the literature
\cite{Cipriani05,Lepri2011,Dhar2013}. This leads naturally to 
consider hydrodynamic equations where the standard Laplacian operator is replaced by
a fractional one \cite{Lepri2010,basile2016thermal,Cividini2017,dhar2019anomalous}.
Nonlinear coupling among hydrodynamic fields  yields slow algebraic decay of current correlations at equilibrium and effective non-local equations. 
A theoretical justification of such a behavior has been obtained by the nonlinear fluctuating hydrodynamics approach, whereby long-wavelength fluctuations are described in terms of the Kardar-Parisi-Zhang equations \cite{Spohn2014}.
Although the resulting predictions are well confirmed by numerical 
simulations~\cite{Mendl2013,Das2014,mendl2015current}, there may be significant scale-effects, especially
when the dynamics is only weakly chaotic \cite{lepri2020too}.

In presence of long-range forces one may expect non-local effective hydrodynamic description to arise naturally by the non-local nature of couplings. 
Indeed, perturbations may propagate with infinite velocities, making this class of systems qualitatively different from their short-ranged counterparts \cite{Torcini1997,Metivier2014}. This has also effects on energy transport 
for open systems interacting with external reservoirs and, more 
generally, on the way in which the long-range terms couple the system
with the environment. For nonlinear oscillator assemblies this problem has
received some attention in the recent literature \cite{avila2015length,Olivares2016,Bagchi2017,bagchi2017energy,Iubini2018,wang2020thermal} 
but is less developed with respect to the short-range case.
A further complication is that relaxation to local and global equilibrium
may occur through  long-living  metastable states and exhibit anomalous diffusion of energy \cite{Bouchet2010,RuffoRev,Campa2014} and even lack of thermalization upon interaction with a single external bath \cite{deBuyl2013}.

Another motivation of the present study is to understand the role of 
phase transitions on anomalous transport in low-dimensional systems. 
As it is known, there is a well-established theory of dynamical critical 
phenomena \cite{hohenberg1977theory} which gives insight on transport coefficients
at criticality. However, studies of energy transport in microscopic
models are scarce, mostly limited to spin systems 
see e.g. \cite{harris1988thermal,saito1999transport,colangeli2018nonequilibrium}
for some related results on critical Ising models with nearest-neighbor
coupling.
 
Besides the above theoretical motivations, the physics of long-range 
interacting oscillator arrays has interesting experimental applications. 
The most natural candidates are trapped ion chains, where 
ions can be confined in periodic arrays and can be studied in an open setup \cite{bermudez2013controlling,ramm2014energy}.
Quantum spins with tunable long-range interactions 
can be realized in the laboratory and the propagation of 
a disturbance can be studied \cite{richerme2014non}. 
On a macroscale, effective long-range forces arise for tailored macroscopic systems like chain of coupled magnets \cite{moleron2019nonlinear} and the effects
of fluctuations and nonlinearity may be relevant.
Several other application have been considered, and we refer
to \cite{defenu2021long} for a recent account.

In this paper, we study the hydrodynamic properties of a one-dimensional  
chain of single or double-well oscillators 
interacting through a long-range potential that can be either 
attractive or repulsive. It can be seen
as a discretization of a $\varphi^4$ field theory, directly related to the long-range 
Ising model and, thus, an effective description of spin chains with 
coupling decaying as an inverse power of their relative distance \cite{richerme2014non}.
The paper is organized as follows. In Section 2 we define the model and discuss its general 
properties. In Section 3 we present the results obtained from equilibrium simulations concerning the dynamical scaling analysis of structure factors and of excess energy correlations. The single-well potential  and the double-well case are separately analyzed. 
In Section 4 we introduce an effective stochastic model of long-range transport based on a L\'evy flight process and we show that it can
reproduce the dynamical scaling observed in the  $\varphi^4$ lattice. 
Stationary out-of-equilibrium states resulting from application of suitable thermal
imbalances and the corresponding scaling of heat fluxes are also analyzed.
Steady nonequilibrium states of the $\varphi^4$ model are presented separately in Section 5 along with a discussion of finite-size effects. 
Finally, Sections 6 and 7 are respectively devoted to a discussion
of the results and to some concluding remarks.

\section{The long-range-interacting $\varphi^4$ lattice in one dimension}

We consider a one-dimensional lattice of $N$ particles with periodic boundary conditions, whose dynamics is governed by the long-range Hamiltonian 
\begin{equation}
\label{eq:H}
H = \sum_{i=1}^N \left[\frac{p_i^2}{2} + U(q_i)-
\frac{\mu}{\mathcal N_\sigma}\sum_{j> i}^{N} \,\frac{q_i q_j}{d_{ij}^{1+\sigma}} \right]
\end{equation}
where $\mu$ is a coupling constant.  The $q_i$s are continuous real variables that denote the oscillator displacements while $p_i=\dot q_i$ are the corresponding momenta
(we set all the masses to unity henceforth). The cases $\mu>0$ and $\mu<0$ correspond to ferromagnetic (attractive) and antiferromagnetic (repulsive) interactions, respectively. 
The quantity $d_{ij}$ identifies the shortest distance between sites $i$ and $j$ on a periodic lattice 
\begin{equation}
\label{eq:d}
d_{ij}=\min\{|i-j|,N-|i-j|\}.
\end{equation}
The real exponent $\sigma\ge-1$ is the parameter that controls the interaction range.
The case $\sigma=0$ identifies the extensivity threshold in one dimension~\cite{RuffoRev}. We follow the usual Kac prescription to keep 
the potential energy extensive even when $\sigma<0$, by defining 
\be
\label{eq:kac}
{\mathcal N}_\sigma  =   2\sum_{r=1}^{N/2}\frac{1}{r^{1+\sigma}}.
\ee
Notice that for $\sigma = -1$, i.e. the case of a mean-field interaction,   
$N_{-1}=N$. For $\sigma > 0$ ${\mathcal N}_\sigma $ attains a constant value for large
sizes $N$ and diverges for $\sigma < 0$.
Finally, in the limit of $\sigma\to+\infty$ the case of nearest-neighbor interactions is retrieved. In the present work we will concentrate on the weak-long range 
case, $\sigma>0$.

For generic choices of the local potential $U(q)$ the model is nonintegrable, with the energy 
being the only constant of motion. 
We consider here the $\varphi^4$ potential  \cite{staniscia2019differences}
\begin{equation}
U(q)= \pm\frac{q^2}{2}+\frac{q^4}{4};
\label{pot}
\end{equation}
in suitable units.  Let us first discuss the attractive coupling $\mu>0$. 
The single-well case, $+1$ in (\ref{pot}), 
identifies  the long-range generalization 
of the non-linearly pinned chain (or discrete Klein-Gordon field). The ground state corresponds
to all particles at rest with $q_i=0$ and the model is expected to admit only a disordered
(paramagnetic) phase. Transport properties in the nearest-neighbor
case have been thoroughly studied in the past \cite{hu2000heat,Aoki00,Piazza2009}.
Generically, energy fluctuations diffuse normally, following the standard
heat equation with a finite thermal diffusivity in the thermodynamic limit.

The double well case, $-1$ in (\ref{pot}), is of particular interest
and indeed its nearest-neighbor version has been considered
starting from the seventies of last century \cite{krumhansl1975dynamics}. It can be 
can be seen as a Hamiltonian, lattice version of the familiar
$\varphi^4$ scalar field with long-range forces.
The model has two degenerate ground states with $q_n=g$ and $g=\pm\sqrt{\mu+1}$, corresponding to an energy density $e=H/N=-(\mu+1)^2/4$ independently on $\sigma$ (notice that the ground state displacement $g$ does not coincide with the minima of $U(q)$,
which are equal to $\pm 1$). 
%it is different from the minima of $U$ which are equal to $\pm 1$).
%\red{mi viene che in (\ref{eq:H}) la somma sul termine LR è $\sum_{j>i}$ non $\sum_{j\neq i}$. Ho verificato che 
%le mie simulazioni sono in accordo con questa scelta.}
%
Since the model has the same symmetries of the ferromagnetic Ising model with 
long-range couplings \cite{dyson1969existence}, its equilibrium
critical properties should be in the same universality class. 
Indeed, one can think of the $q_i$s as a
coarse-grained version of the Ising spins. It is thus natural to consider 
as order
parameter the average of the ``magnetization'' density
\[
m =  \frac{1}{N}  \sum_{i=1}^N q_i 
\]
Remarkably, depending on $\sigma$ the double-well chain can display a phase 
transition in the weak-long range regime. This can be justified as follows. 
Following \cite{mukamel2009notes}, for weak long
range interactions in $d$ dimensions one can write a Landau-Ginzburg effective
free energy of the form
\begin{equation}
F_{GL}= \int d^dr\left[\frac{1}{2}a_2m^2({\bf r}) + \frac{1}{4}a_4 m^4({\bf r})\right] + \int d^dr ~
d^dr' \frac{m({\bf r})m({\bf r'})}{|{\bf r-r'}|^{d+\sigma}} ~,
\end{equation}
where $a_2$, $a_4$ are suitable  coefficients and $m(\bf r)$ is the magnetization field at point ${\bf r}\in\mathbb{R}^d$. The second term 
represents the contribution of the long-range
forces to the energy. In terms of the Fourier components of the
order parameter $m({\bf k})=\int d^d r\, m({\bf r}) e^{i {\bf k} \cdot \bf r} $ this integral may be expressed as a
Gaussian approximation (far from criticality).
Accordingly, to leading order in $k$, the Fourier transform of the long range
potential is of the form $a+bk^\sigma$, where $a$ and $b$ are
constants. Therefore one obtains
\begin{equation}
\label{HLongRange} F_{GL} = \frac{1}{2V}\sum_{{\bf k}}(\bar{a}_2+
bk^\sigma+k^2) m({\bf k})m(-{{\bf k}}) ~,
\end{equation}
where $\bar{a}_2=a_2+a$. The $k^2$ term results from short range
interactions which are always present in the system.
For $\sigma > 2$ the $k^\sigma$ term in (\ref{HLongRange}) is dominated by the
$k^2$ term in the long-wavelength limit and may thus be neglected. One is then back to the model
corresponding to short range interactions and the upper critical
dimension is $d_c=4$. On the other hand, for $0<\sigma<2$ the
term is $k^\sigma$ dominates and the correlation length
diverges as $\xi\propto|\bar{a}_2|^{-1/\sigma}$ \cite{mukamel2009notes}.
%%
%%
%\begin{equation}
%\xi~ \propto ~|\bar{t}|^{-1/\sigma}~.
%\end{equation}

This expectation has been demonstrated numerically and analytically 
for the mean-field case 
$\sigma=-1$ both in the canonical \cite{desai1978statistical}
and microcanonical \cite{dauxois2003clustering} ensembles.
In the range $0<\sigma<1$, the transition is of second order,
separating a ferromagnetic phase at low temperatures from
a paramagnetic phase at high temperatures. 
For $\sigma>1$, the system is disordered at all energies. The
case $\sigma=1$  is peculiar, since it shows a Kosterlitz-Thouless
phase transition with a discontinuous jump in the magnetization \cite{aizenman1988discontinuity}. 
For $1<\sigma<d/2$ the transition is mean-field with exponents
independent on $\sigma$, while for $d/2<\sigma<1$ 
they depend continuously on $\sigma$ \cite{mukamel2009notes}.
We refer also to  \cite{staniscia2019differences}
for further studies of critical properties of the  model .

The antiferromagnetic,  repulsive case $\mu<0$ has been studied 
in the mean-field limit $\sigma=-1$, where it was shown that no 
phase transition occurs
in this limit \cite{dauxois2003clustering}.
To our knowledge,  the case  $\sigma>-1$ 
has not been studied 
so far.  Based again on the analogy with the antiferromagnetic Ising model
with long-range interaction \cite{kerimov1993absence}
we do not expect any phase transition here.

In the following Section we will present some numerical results
that confirm the above expectations.

\section{Equilibrium simulations}\label{sec:eq.}

In this Section we report the results of equilibrium  microcanonical simulations of 
the model. Numerical integration of the associated Hamilton equations 
can be computationally demanding in long-range interacting systems.
However, since the distance $d_{ij}$ only depends on $|i-j|$, it turns out that the
%the coupling matrix among  oscillators
matrix of coupling strengths among oscillators, namely $d_{ij}^{-1}$, is circulant.
One can thus exploit translational invariance to compute the  
forces as convolution products 
by an algorithm based on the Fast Fourier Transform \cite{Gupta2014}.  
The simulation protocol to sample an equilibrium configuration for a given temperature $T$ is the following.
\begin{itemize}
\item For the double-well local potential,
canonical variables are initialized as
 $q_n(0)=\pm 1$ with equal probability 
and $p_n(0)$ randomly drawn from a Gaussian distribution  with width $\sqrt{T}$. For the single-well potential,
initial displacements are chosen as $q_n(0)=0$;
\item The system is let evolve for a canonical transient whereby 
a subset of randomly chosen particles (typically $5\%$) undergo random collisions with 
a Maxwellian bath at temperature $T$. Each collision event amounts to assign 
to the colliding particle a new velocity $p_n = v$, with $v$ being a random Gaussian
with width $\sqrt{T}$. This allows to thermalize the initial 
configuration and fix the energy density $e=H/N$ corresponding to the desired temperature;
\item Finally, Maxwellian collisions are switched off and a microcanonical trajectory is generated by means of a 4th-order symplectic algorithm \cite{mclachlan1992accuracy}. Statistical averages of 
the relevant observables,  as the kinetic temperature 
$T=\langle p_i^2\rangle$, are thereby computed.
%\item observables: magnetization ad susceptibility
%\[
%M = \overline{\frac{1}{N} \sum_n q_n}\qquad  
%\chi= \overline{\left(\frac{1}{N} \sum_n q_n\right)^2}-\overline{\frac{1}{N} \sum_n q_n} ^2
%\]
\end{itemize}

\begin{figure}
\hfil
\includegraphics[width=0.9\textwidth]{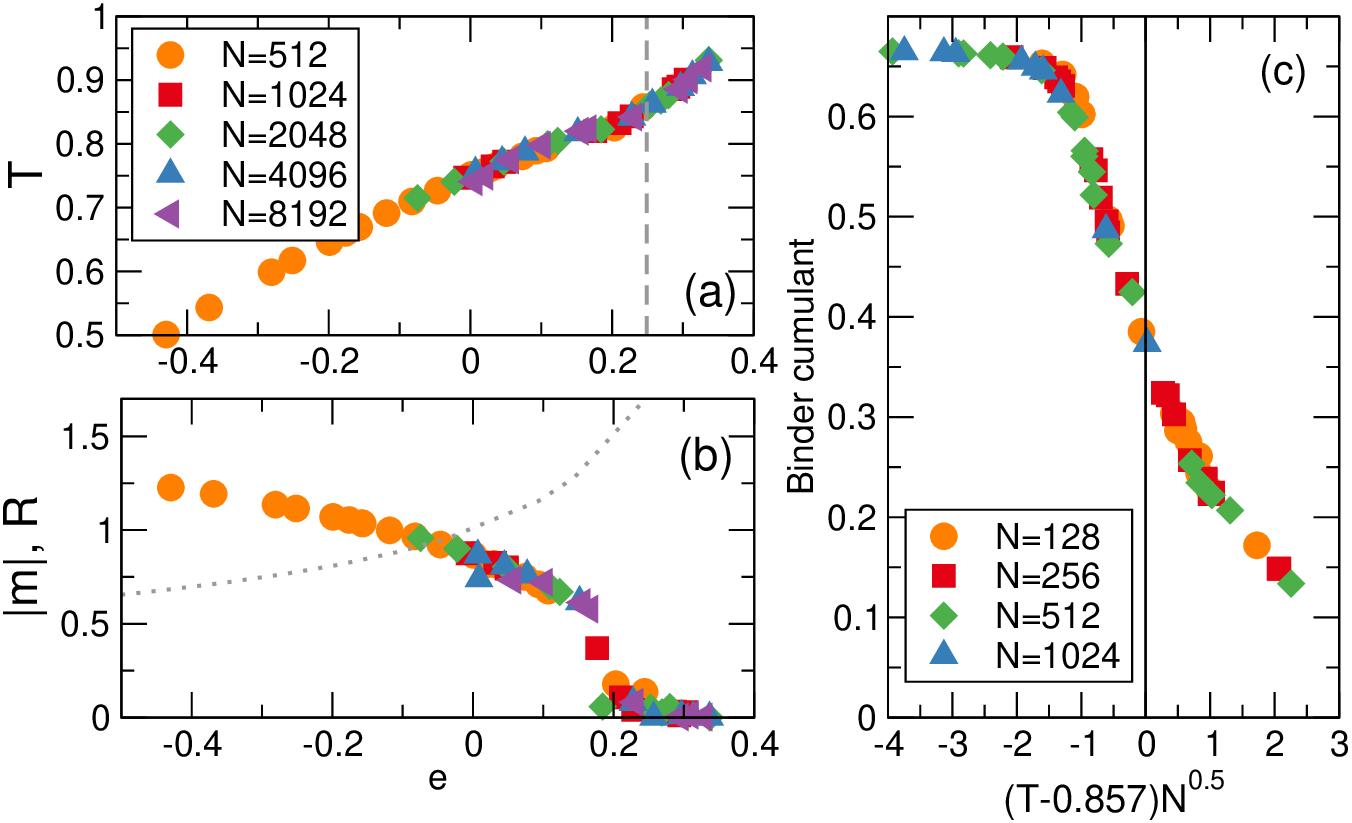}
\caption{Illustrating the second-order phase transition in the double-well model, 
$\mu=1$, $\sigma=0.6$, microcanonical simulation as described in the main text: 
(a) caloric curves for different lattice
sizes and (b) absolute value of the magnetization 
(symbols); the dotted line is the ratio $R$ between the 
average of the anharmonic and harmonic energies (see text). Panel   
(c) illustrates the finite-size scaling analysis of the Binder cumulant $B$ as a function 
of the scaled temperature $(T-T_c)N^{0.5}$.  $T_c$ is estimated by looking for the
intersection point of the curves $B(T,N)$. The vertical line in (a)  signals the 
best estimate of the critical energy
$e_c\approx 0.249$ corresponding to $T_c\approx 0.857$.}
\label{fig:ptrans}
\end{figure}

For the sake of illustration, we report in fig.\ref{fig:ptrans} the 
magnetization and caloric curves for different lattice  
sizes, as obtained by microcanonical simulation
of the double-well ferromagnetic ($\mu>0$) model. Up to the statistical fluctuations and finite-size effects, a continuous transition 
in $m(e)$ at 
a critical energy density $e_c$ is clearly seen, along with 
a discontinuos change of the slope of  the caloric curve $T(e)$
at $T_c=T(e_c)$. 
The difference between the two phases can be 
seen by computing the
equilibrium probability distribution of the $q_i$'s (data not shown). In the sub-critical phase $T<T_c$ this distribution
is asymmetric with a main peak in correspondence of one of the two degenerate ground states, while it is symmetric in the disordered
phase $T>T_c$, as expected. To quantify the strength of the 
nonlinear term we also report in figure~\ref{fig:ptrans}(b) the  
ratio $R=\langle H_4\rangle/\langle H_2\rangle$ where 
$H_2$ and $H_4$ are, respectively, the quadratic and quartic parts of the 
Hamiltonian (\ref{eq:H}). Since $R$ is of order one in the 
considered energy regimes, this clearly confirms that we are very far 
from the weakly-nonlinear case.

To locate the critical point we used the well-known finite-size scaling method, 
(see e.g. \cite{caiani1998hamiltonian} and references therein 
for application to the short-ranged
$\varphi^4$ lattice).  We computed 
the Binder cumulant $B(T,N)=1-\langle m^4\rangle/3\langle m^2\rangle^2$
for different lattice sizes. Following the standard prescription this alllows to
 estimate the critical kinetic temperature $T_c$, as shown in fig.\ref{fig:ptrans}c.
The accuracy of such  estimates are  sufficient for our purposes and 
we did not attempt to measure critical exponents.

We also performed some simulations for the double-well model with 
repulsive interaction, setting $\mu=-1$.   For the explored cases, 
the average magnetization $\langle m\rangle$ is vanishingly small
and the caloric curves do not display any sign of discontinuities. 
This confirm the expectation that the repulsive model is not critical.

Since we are interested in large scale (hydrodynamic) properties, we investigated
equilibrium correlation functions in the long-time and distance regime.
As it is known, this gives direct information on 
propagating modes on such scales. In heat 
transport problems one is primarily interested in the propagation of 
energy fluctuations. We thus consider the site energies
\begin{equation}
h_i = \frac{p_i^2}{2} + U(q_i)-
\frac{\mu}{{\mathcal N}_\sigma }\sum_{j\neq i}^{N} \,\frac{q_i q_j}{d_{ij}^{1+\sigma}}, 
\end{equation}
which are inherently non-local.  Moreover, despite our study is focused on 
energy transport ad relaxation, in the following we will 
also report some correlation data of the $q_n$, namely of the local
magnetization.

For computational reasons it is convenient to work in the Fourier domain,
i.e. we first perform the discrete transform
\begin{equation}
h(k,t)  = {1\over N}\,\sum_{n=1}^N \, h_n \exp(-ik n), \quad 
m(k,t)  = {1\over N}\,\sum_{n=1}^N \, q_n \exp(-ik n)
\label{dens}
\end{equation}
By virtue of the periodic boundaries, the allowed
values of the  wavenumber $k$ are integer multiples of $2\pi/N$.
We then define the dynamical structure factors, namely the square modulus of 
the temporal Fourier transform,  as
\begin{equation}
S_h(k,\omega) \;=\; 
\big\langle \big| h(k,\omega )\big|^2 \big\rangle,  \quad 
S_m(k,\omega) \;=\; 
\big\langle \big| m(k,\omega )\big|^2 \big\rangle  \quad .
\label{strutf}
\end{equation}
The square brackets denote an average over a set of independent
molecular--dynamics runs (typically a few hundreds). 
Simulations have been performed for 
a relatively large lattice size $N=2048$. In some case a check of 
finite-size effect has been done by comparing the structure 
factors (at fixed $k$) for $N=4096$.

A complementary approach is to compute the spatio-temporal correlations, that are useful to ascertain the 
nature of the energy diffusion process both in short \cite{Zhao06,Li2015} 
and long-range \cite{Bagchi2017} interacting models.
Using the standard prescriptions, we partition the lattice in $N_B$ boxes 
and coarse-grain the local energy 
fluctuations $\Delta h_i=h_i-e$ to compute 
\begin{equation}
C(x,t) = 
\frac{\langle \Delta h_i(t)\Delta h_j(0)\rangle  }{\langle \Delta h_i(0)\Delta h_i(0)\rangle}
+\frac{1}{N_B-1}
\end{equation}
averaged over
the microcanonical ensemble (this is sometimes referred to as 
excess energy correlation \cite{Zhao06,Li2015,Bagchi2017}). As usual, translational invariance
is assumed, making $C(x,t)$ depend only on the relative distance $x=|i-j|$.

%Dynamical scaling means that for $q,\omega \to 0$
%\begin{equation}
%S_h(q,\omega) \;\propto\; \varphi\left( \frac{\omega}{q^{z_h}} \right)  
%\quad S_m(q,\omega) \;\propto\; \psi\left( \frac{\omega}{q^{z_m}} \right)  
%\end{equation} 
%the dynamical exponent $z$ is 2 for normal diffusion.

\subsection{Single-well potential}

Let us first discuss  the case of the single-well model. Here the 
average magnetization is vanishing. Since energy is the only conserved 
quantity, we can restrict ourselves to the analysis of 
% can consider only
  its fluctuations as measured by
$S_h$.

\begin{figure}
\hfill
\includegraphics[width=0.9\textwidth]{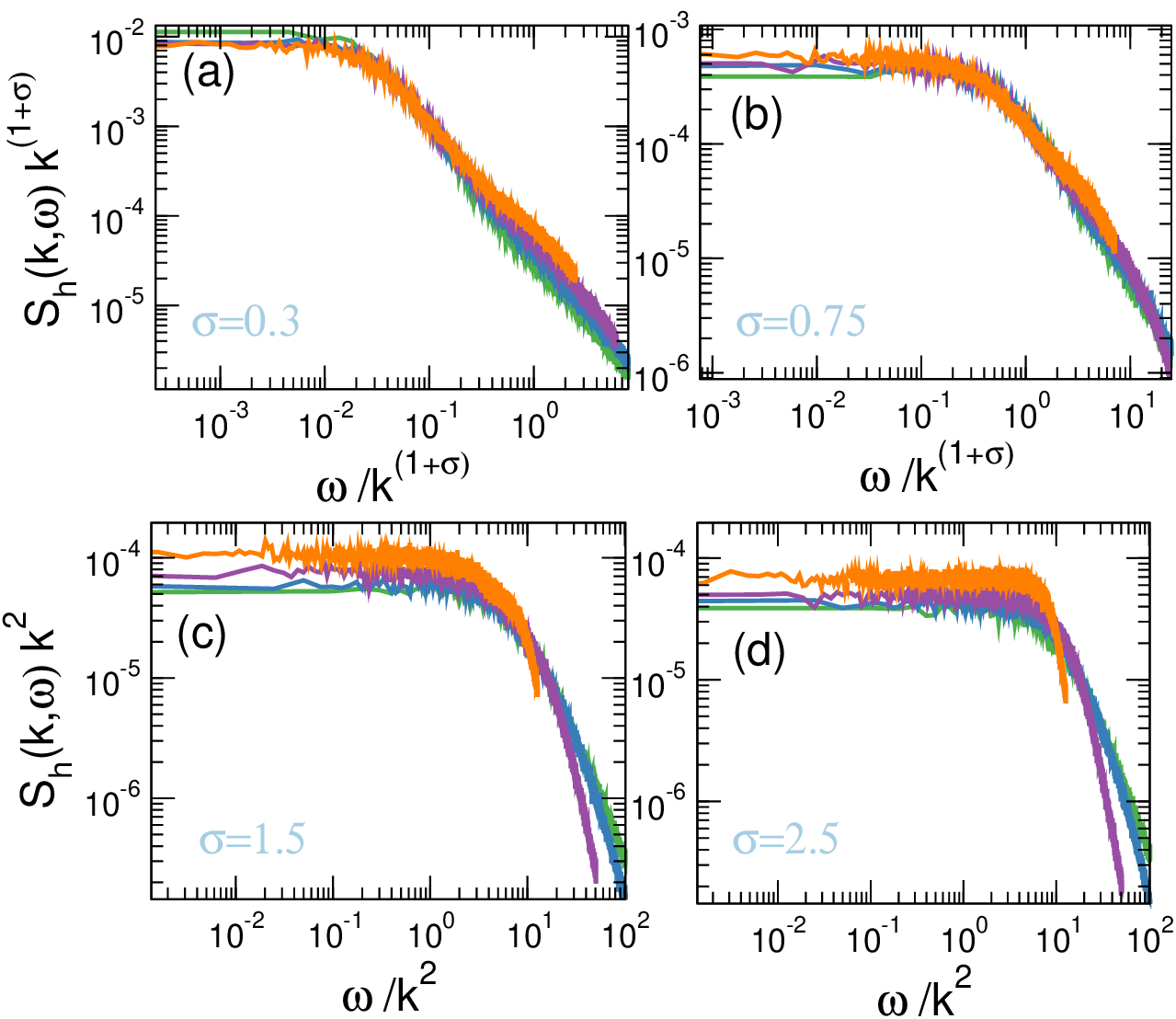}
\caption{Energy structure factors for the single-well potential
case for various values of the range exponent $\sigma$ and different 
wavenumbers $k=2\pi l/N$, $l=4,8,16,32$s.
Simulations with $T=0.2$, $N=2048$ and $\mu=1$.}
\label{fig:sw}
\end{figure}

The data collapse in fig. \ref{fig:sw}, obtained for $T=0.2$, shows that in the hydrodynamic limit 
$k,\omega \to 0$ there is a well-defined dynamical scaling
\begin{equation}
S_h(k,\omega) \;=\; \frac{1}{k^{z_h}}\Phi\left( \frac{\omega}{k^{z_h}} \right)  
%\quad S_m(q,\omega) \;\propto\; \Psi\left( \frac{\omega}{q^{z_m}} \right)  
\label{eq:scal}
\end{equation} 
where the dynamical exponent $z_h$ is compatible with the 
values 
\begin{equation}
z_h=1+\sigma \quad \mathrm{for} \;0<\sigma<1;\qquad 
z_h=2  \quad \mathrm{for}\; \sigma>1.
\label{eq:zh}
\end{equation}
Note that rescaling of data in fig. \ref{fig:sw} is parameter-free.
Moreover, the scaling function  can be rather accurately fitted with a 
Lorenzian lineshape $\Phi(x)=a/(1+bx^2)$, with $a,b$ being 
fitting parameters.
This result appears to be independent of the energy density 
or temperature (at least for $e$ not too small). A series of 
simulations with $T=2$ (not reported) yields the same scaling.

\begin{figure}
\hfill \includegraphics[width=0.9\textwidth]{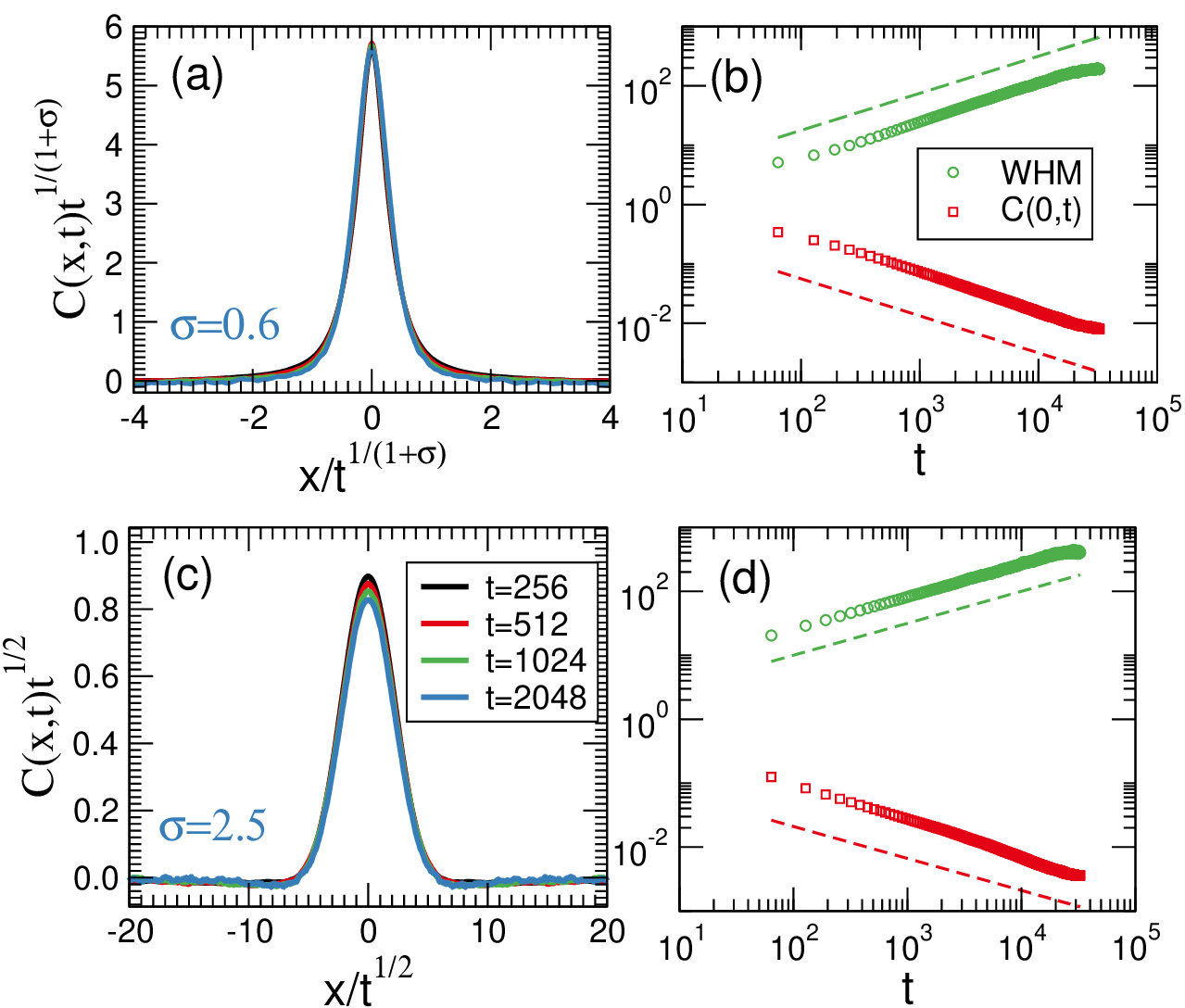}
\caption{Excess energy correlations $C(x,t)$ for the single-well potential,
$\mu=1$, $N=2048$, $T=0.2$
(a,b) $\sigma=0.6$, (c,d) $\sigma=2.5$; 
Left panels: rescaled correlation, Right panels: the time dependence of the width at half 
maximum (WHM) and the maxima $C(0,t)$; the dashed 
lines indicate the power-law scalings $t^{\pm 1/(1+\sigma)}$ (c)
and $t^{\pm \frac{1}{2}}$ (d)  corresponding to L\'evy and 
normal diffusion
respectively. Energy fluctuation are coarse-grained over 4 adjacent 
lattice sites.}
\label{fig:swexcess}
\end{figure}

To further confirm the above scenario, we have performed a series of simulations
of the excess energy correlations.
Figure \ref{fig:swexcess} reports
$C(x,t)$ at different times, rescaled as $t^{1/z_h}C(xt^{-1/z_h},t)$ where is
$z_h$ the  dynamical exponent given above. The data collapse (fig. \ref{fig:swexcess}a,c) 
is excellent and both the width at half-maximum of $C(x,t)$ as well as the maximum of the pulses $C(0,t)$ follow well the expected power laws over more 
than two decades (fig. \ref{fig:swexcess}b,d). 
Actually, there are some deviations at small $x$ that however
tend to reduce upon increasing $t$.
%Moreover, we show in figure~\ref{fig:excess3} analogous data of $C(x,t)$ for the case 
%of repulsive interactions, $\mu<0$. The results are qualitatively very similar to 
%the attractive case and the scaling laws are the same. Thus, we can argue that hydrodynamics is not affected by the sign of the long-range potential.
 %
 %
% \begin{figure}
%\hfill \includegraphics[width=0.9\textwidth]{xcoraf.eps}
%\caption{Excess energy correlations $C(x,t)$ for the single-well potential,
%$\mu=-11$, $N=2048$, $T=0.2$
%(a,b) $\sigma=0.6$, (c,d) $\sigma=2.5$; 
%Left panels: rescaled correlation, Right panels: time dependence of the width at half 
%maximum (WHM) and the maxima $C(0,t)$; the dashed 
%lines indicate the power-law scalings $t^{\pm 1/(1+\sigma)}$ (c)
%and $t^{\pm \frac{1}{2}}$ (d)  corresponding to L\'evy and 
%normal diffusion
%respectively. Energy fluctuation are coarse-grained over 4 adjacent 
%lattice sites.}
%\label{fig:excess3}
%\end{figure} 

According to the above results, for $\sigma>1$ energy fluctuations obey standard diffusive scaling, 
as it is well-known for the short-range case. This result matches well with
the general expectation that in this regime the long-range terms do
not affect the large-scale dynamics and that transport is normal.
Conversely, the weak-long range case $0<\sigma<1$ shows an anomalous scaling with 
a $\sigma$-dependent dynamical exponent.  

We can rationalize the results for $0<\sigma<1$ by assuming that fluctuations of 
the energy field $h(x,t)$ obey the stochastic hydrodynamics equations
\begin{eqnarray}
&\dot h =-\partial_x J\nonumber\\
&{J} = -  \partial_x^\sigma 
\left( D_\sigma h + \eta \right)
\label{eq:fdif}
\end{eqnarray}
where $\eta(x,t)$ is a standard space-time	 noise with correlations
$\langle \eta(x,t)\eta(x',t')\rangle =R\delta(t-t')\delta(x-x')$, 
$\partial_x^\sigma$ is a shorthand notation for the fractional 
derivative of order $\sigma$ (we assume the standard Riesz definition 
in terms of the Fourier transform)  and $D_\sigma$ represents 
a generalized diffusion coefficient. 
The first equation in (\ref{eq:fdif})
is the continuity equation defining the energy current $J$, 
while the second can be considered as a generalization of 
the Fourier law to include long-range transport,
namely the fact that energy flow is inherently non-local due to nature of the the coupling.

Since (\ref{eq:fdif}) is linear we can straightforwardly compute
the structure factor by Fourier transform, yielding:
\begin{equation}
S_h(k,\omega) \;=\; R \frac{D_\sigma |k|^{1+\sigma}}
{\omega^2+(D_\sigma |k|^{1+\sigma})^2}
\label{ew:lorenz}
\end{equation}
which clearly satisfies (\ref{eq:scal}).

Altogether, we conclude that stochastic fractional diffusion equation 
(\ref{eq:fdif}) is an accurate description of the microscopic model.
In Section~\ref{sec:levy} we will provide an effective stochastic model that 
accounts for the observations.

\subsection{Double-well potential}

\begin{figure}
\hfill \includegraphics[width=0.9\textwidth]{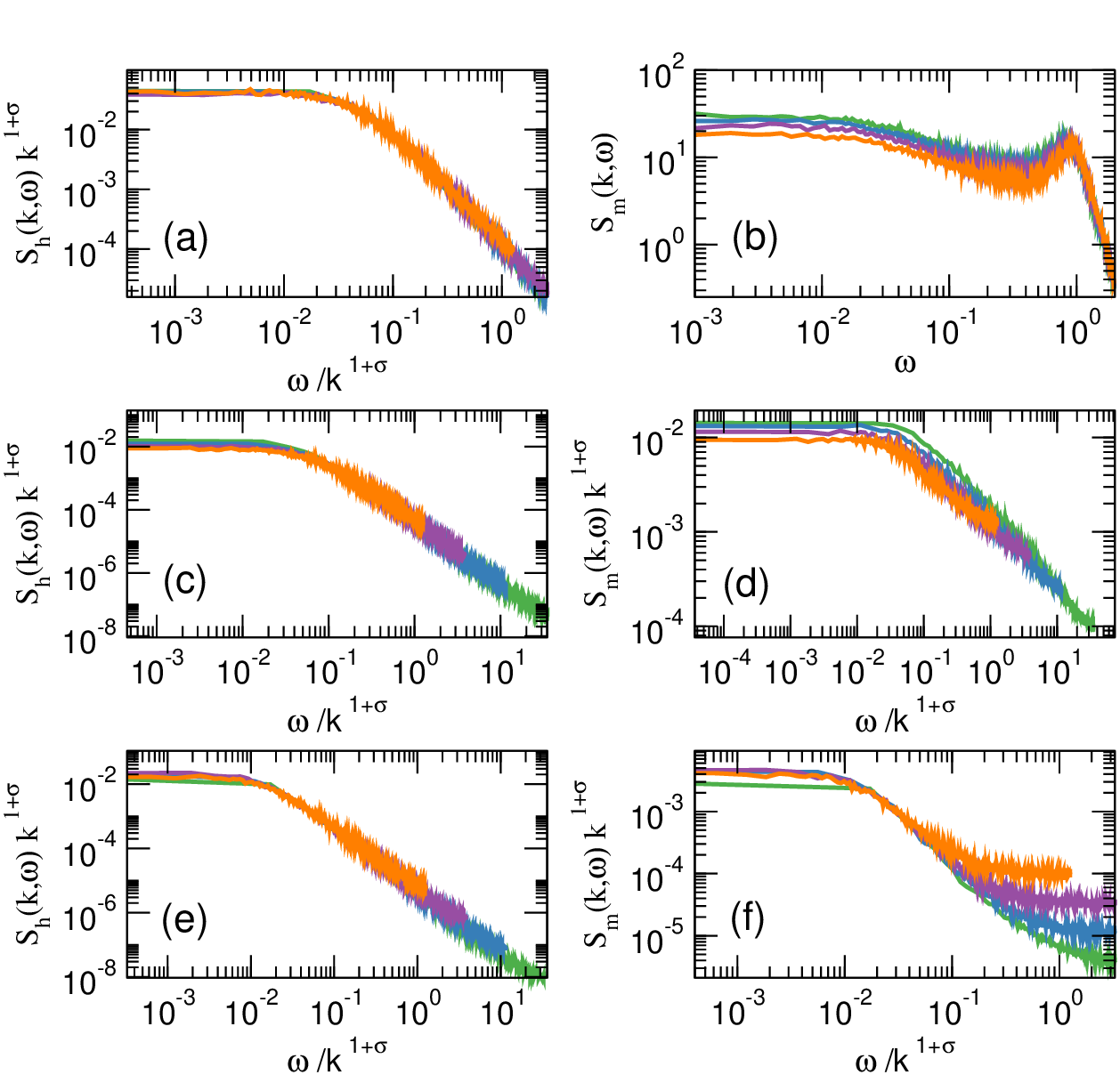}
\caption{Energy (left) and magnetization (right) structure factors for the double-well potential
case for $\mu=1$ $\sigma=0.6$, $N=2048$, $k=2\pi l/N$, $l=4,8,16,32$. Parameters in   panels (a,b), (c,d) and (e,f)
are for supercritical (paramagnetic) $T=2, e=1.13$, close to critical $T=0.85, e=0.249$, and subcritical $T=0.5, e=-0.42$
(ferromagnetic) regions, respectively.}
\label{fig:dw}
\end{figure}

\begin{figure}
\hfill\includegraphics[width=0.9\textwidth]{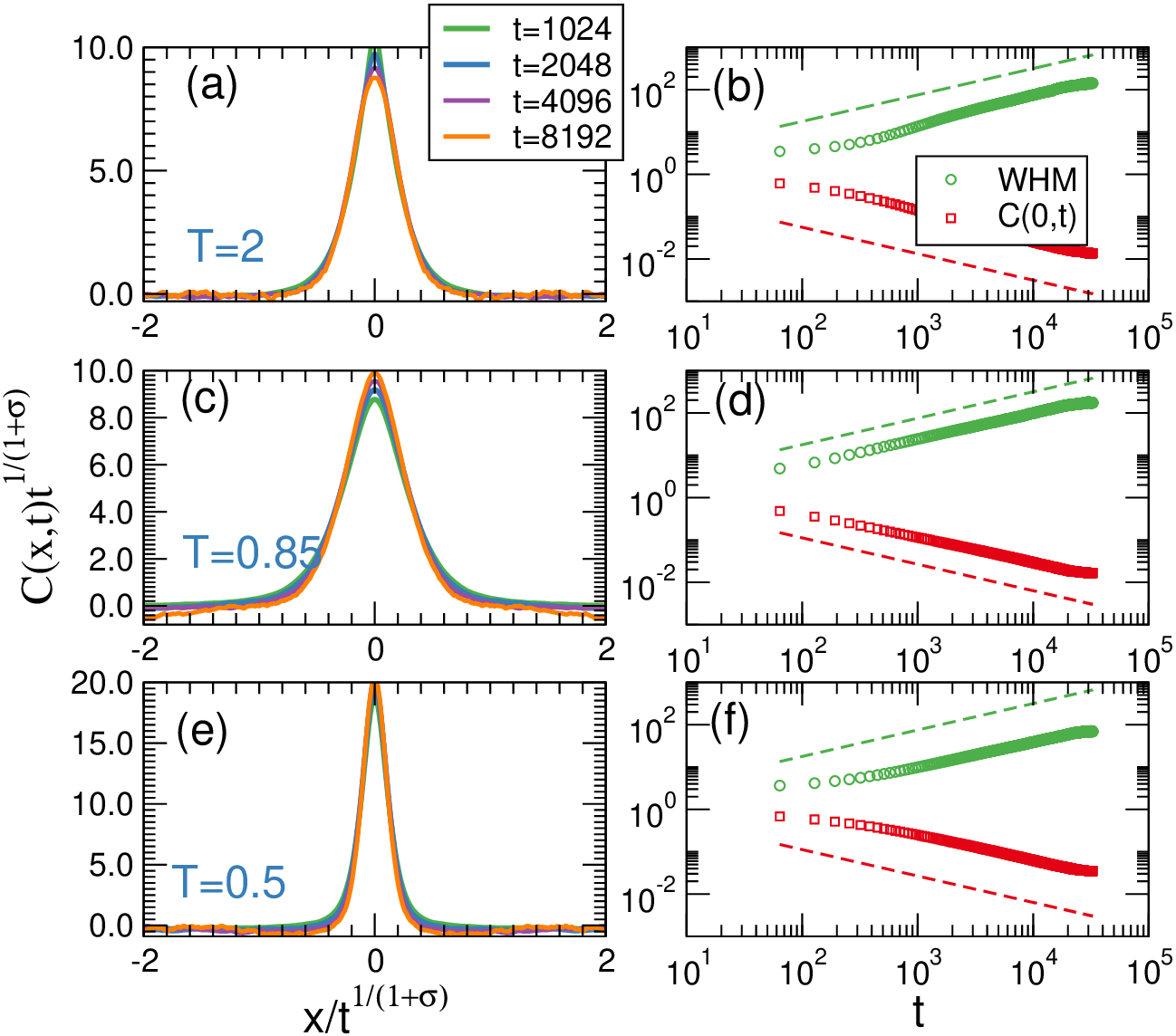}
\caption{Excess energy correlations $C(x,t)$ for the double-well potential 
$\mu=1$, $\sigma=0.6$, $N=2048$; Left panels: rescaled correlation; 
Right panels: time dependence of the width at half 
maximum (WHM) and of the maxima $C(0,t)$.
(a,b) $T=2$, $e=1.13 > e_c$,
(c,d) $T=0.85$, $e=0.249 \approx e_c$,%should be the same even in this case.
(e,f) $T=0.5$, $e=-0.42 < e_c$,
the dotted lines indicate the two scalings 
$t^{\pm 1/(1+\sigma)}$ that correspond to L\'evy diffusion. Energy fluctuation are coarse-grained over 4 adjacent 
lattice sites.}
\label{fig:excess}
\end{figure}

\begin{figure}
\hfill\includegraphics[width=0.9\textwidth]{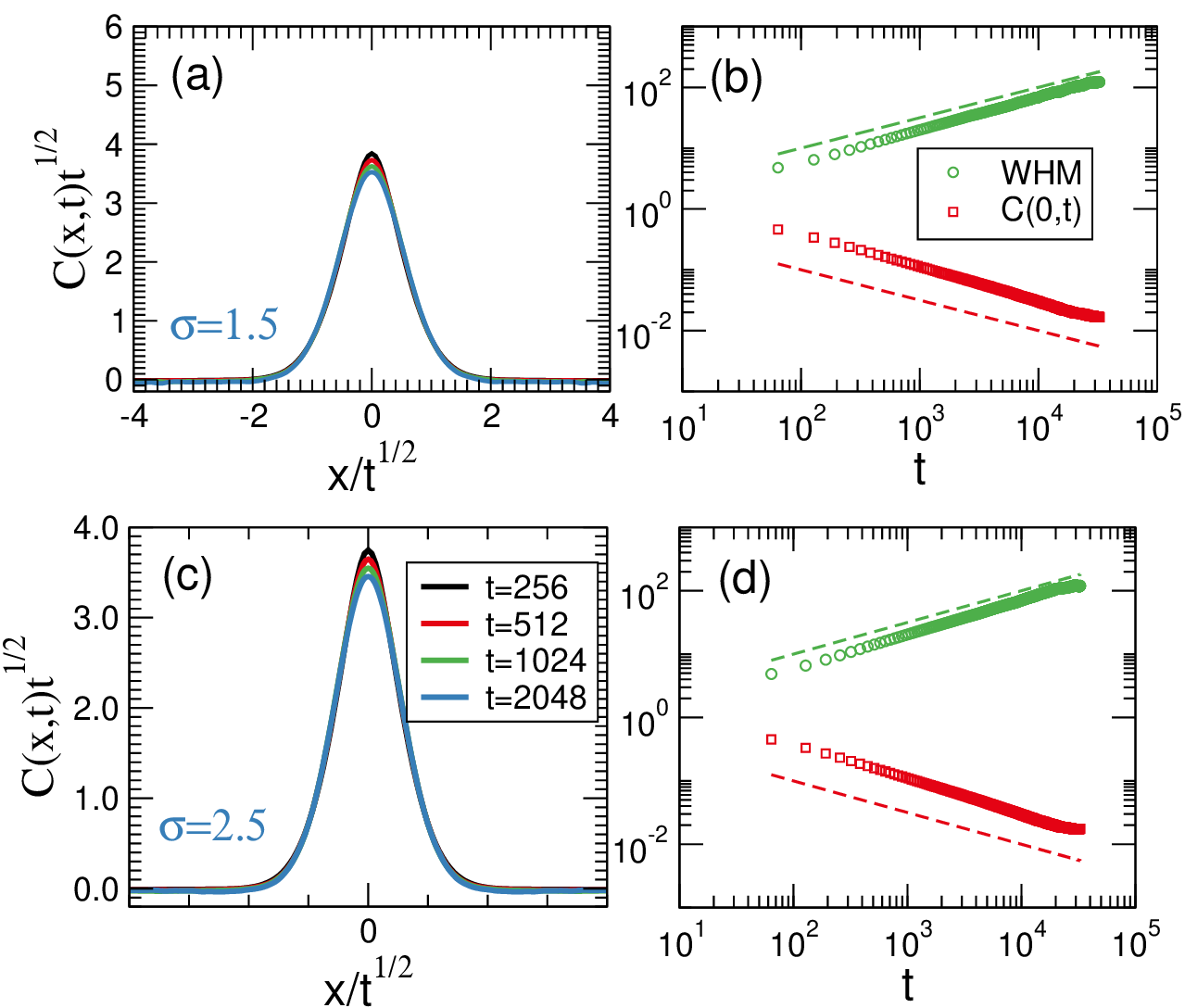}
\caption{Excess energy correlations $C(x,t)$ for the double-well potential 
$\mu=1$,  $N=2048$, $T=2$; Left panels: rescaled correlation; 
Right panels: time dependence of the width at half 
maximum (WHM) and of the maxima $C(0,t)$.
(a,b) $\sigma=1.5$,
(c,d)  $\sigma=2.5$
the dotted lines indicate the two scalings 
$t^{\pm 1/2}$ that correspond to normal diffusion. Energy fluctuation are coarse-grained over 4 adjacent 
lattice sites.}
\label{fig:excess2}
\end{figure}

We now turn to the case of the double-well potential with $\mu>0$. 
The scaled energy structure factors $S_h$ and excess energy correlations $C(x,t)$
for $\sigma=0.6$
are reported in the leftmost panels of figures \ref{fig:dw} and \ref{fig:excess}, respectively 
for three different energies
belonging to the supercritical, critical and subcritical phases.
Figures \ref{fig:excess2} are for $\sigma>1$ where there is no transition.
The data are reported under the same rescalings as in the single-well
case. They confirm that the dynamical scaling of is the same: relation (\ref{eq:scal}) holds with  
$z_h$ as given in (\ref{eq:zh}), for all the considered energies. Remarkably, this remains true even close to the critical point
(see fig. \ref{fig:dw}c and \ref{fig:excess}c).

If a critical point exists, one should consider also the 
dynamics of the magnetization \cite{hohenberg1977theory} and test whether a dynamical scaling 
akin to (\ref{eq:scal}) holds for $k,\omega \to 0$, i.e.
\begin{equation}
%S_h(q,\omega) \;\propto\; \varphi\left( \frac{\omega}{q^{z_h}} \right)  \quad 
S_m(k,\omega) \;=\;\frac{1}{k^{z_m}} \Psi\left( \frac{\omega}{k^{z_m}} \right). 
\label{eq:scalm} 
\end{equation} 
As above, $z_m=2$ for normal diffusion.

For the magnetization correlator the scaling depends on the energy, see the rightmost panels of figure~\ref{fig:dw}.
In the supercritical region, the spectral width of $S_m$ is roughly independent
of the wavenumber (notice that in the corresponding panel of figure \ref{fig:dw}b
frequencies are not scaled). In the subcritical region $S_m$ 
has instead a large low-frequency component that indeed satisfies relation
(\ref{eq:scalm}) with the same exponent as 
$S_h$, i.e. $z_m=z_h=1+\sigma$ for $\sigma <1$.
In the critical region the low-frequency component is still present 
but the quality of data collapse is worse.
%should be the same even in this case.

\begin{figure}
\hfill \includegraphics[width=0.9\textwidth]{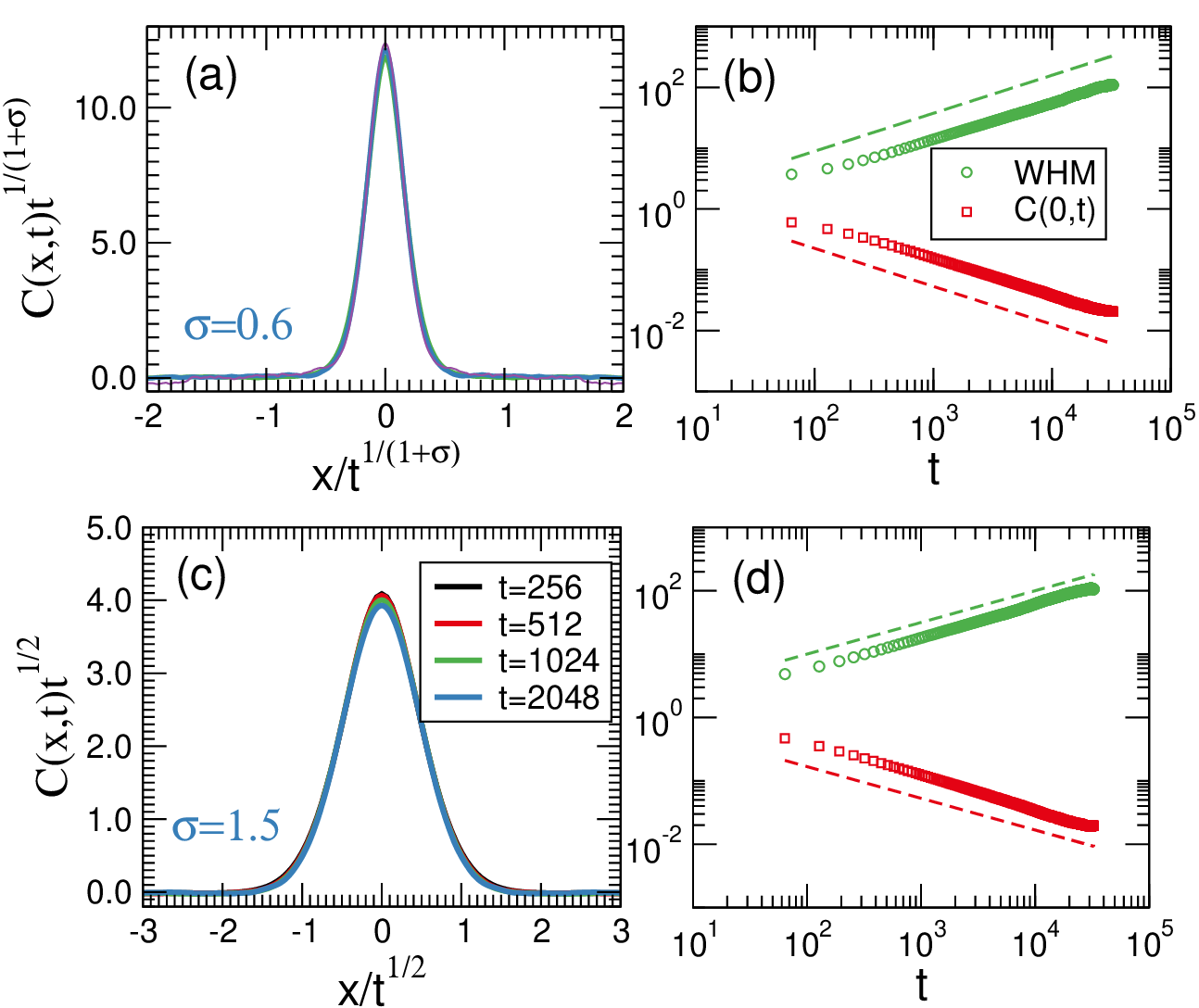}
\caption{Excess energy correlations $C(x,t)$ for the double-well potential
with antiferromagnetic repulsive interaction
$\mu=-1$, $N=2048$, $T=2$
(a,b) $\sigma=0.6$, (c,d) $\sigma=1.5$; 
Left panels: rescaled correlation, Right panels: time dependence of the width at half 
maximum (WHM) and the maxima $C(0,t)$; the dashed 
lines indicate the power-law scalings $t^{\pm 1/(1+\sigma)}$ (c)
and $t^{\pm \frac{1}{2}}$ (d)  corresponding to L\'evy and 
normal diffusion
respectively. Energy fluctuation are coarse-grained over 4 adjacent 
lattice sites.}
\label{fig:excess3}
\end{figure}

Finally, we show in figure~\ref{fig:excess3} the data for the case 
of repulsive interactions, $\mu<0$. The results are the same as 
in the attractive case. We thus conclude that hydrodynamics 
should be the same even in this case.

\section{Effective L\'evy flight model}
\label{sec:levy}

The L\'evy flight process is well known as the simplest generalization
of the Brownian random walk yielding anomalous diffusion 
of an individual particle \cite{metzler2000random,AnotransBook08}.
This idea has been recently proposed to describe hydrodynamic
fluctuations of a quantum spin chain in the infinite-temperature limit \cite{Schukert2020}. 
Let us consider a model where the site energies $h_j$ undergo a 
process in which they are redistributed according to a L\'evy flight 
process, namely
a random walk with step sizes drawn from a distribution with algebraic tails.
The associated master equation reads
\begin{equation}
\dot h_j = \sum_{i\neq j} W_{ij}(h_i-h_j),
\label{eq:ME}
\end{equation}
with transition rates
\begin{equation} 
W_{ij}=\frac{\lambda}{|i-j|^{\nu}}.
\end{equation}
where $\lambda$ is a characteristic rate, setting the inverse timescale of the 
process and $\nu$ is a positive exponent.

In absence of external sources, 
the dynamical correlation of the model can be computed by Fourier transforming
the master equation (\ref{eq:ME}) and considering the large-wavelength limit 
$k\to 0$ of the field $h(k,t)$.
The long-time asymptotics is ruled by long-wavelengths fluctuations.
For $1<\nu<3$, taking into account the leading terms in $k$ one obtains 
(see e.g.  \cite{Schukert2020})
\begin{equation}
\dot h(k,t) \approx -\lambda c_\nu\,|k|^{\nu-1} h(k,t)
%\quad\Rightarrow\quad f(k,t)=f(k,0)\,e^{-\lambda c_\nu\,|k|^{\nu-1} t}.
\label{eq:hdot}
\end{equation} 
where $c_\nu>0$ is a suitable positive constant.
Assuming a localized initial state e.g. $h_j(t=0)\propto\delta_{j,0}$, then 
$h(k,0)$ is a constant. Equation (\ref{eq:hdot}) can be solved 
straightforwardly by Laplace transform, yielding 
\begin{equation}
|h(k,\omega)|^2 \propto 
\frac{ \lambda c_\nu |k|^{\nu-1}}{\omega^2+( \lambda c_\nu |k|^{\nu-1})^2}
\end{equation}
This expression coincides with the structure factor given by (\ref{ew:lorenz})
upon letting 
\begin{equation}
\nu=\sigma+2
\label{es:sigma2}
\end{equation}
and $D_\sigma=\lambda c_\nu$.
 
To further validate the stochastic model, we now consider the out-of-equilibrium
setup where an ensemble of L\'evy fliers is in contact with two reservoirs \cite{Lepri2011,Dhar2013}. To model this situation, we refer to
a finite lattice of $N$ sites, assuming periodic boundary conditions. 
We add to the right hand side of (\ref{eq:ME}) the source 
terms 
$$
s_j=-\gamma \left[\delta_{j,1}(h_j-\varepsilon) + \delta_{j,N/2}(h_j+\varepsilon)\right],
$$
which tend to force the values of $h_j$ to $\pm \varepsilon$ at sites $1,N/2$ 
respectively, with a rate constant $\gamma$. 
 The resulting master 
equation is a linear problem that can be solved numerically 
in the steady state, $\dot h_j=0$. We compute in particular,
$\gamma(h_1-\varepsilon)=Q$ and $\gamma(h_{N/2}+\varepsilon)=-Q$ which are 
the fluxes exchanged with the sources, and the local energy 
profiles $h_j$ in the steady state. 

\begin{figure}
\hfill
\includegraphics[width=0.9\textwidth]{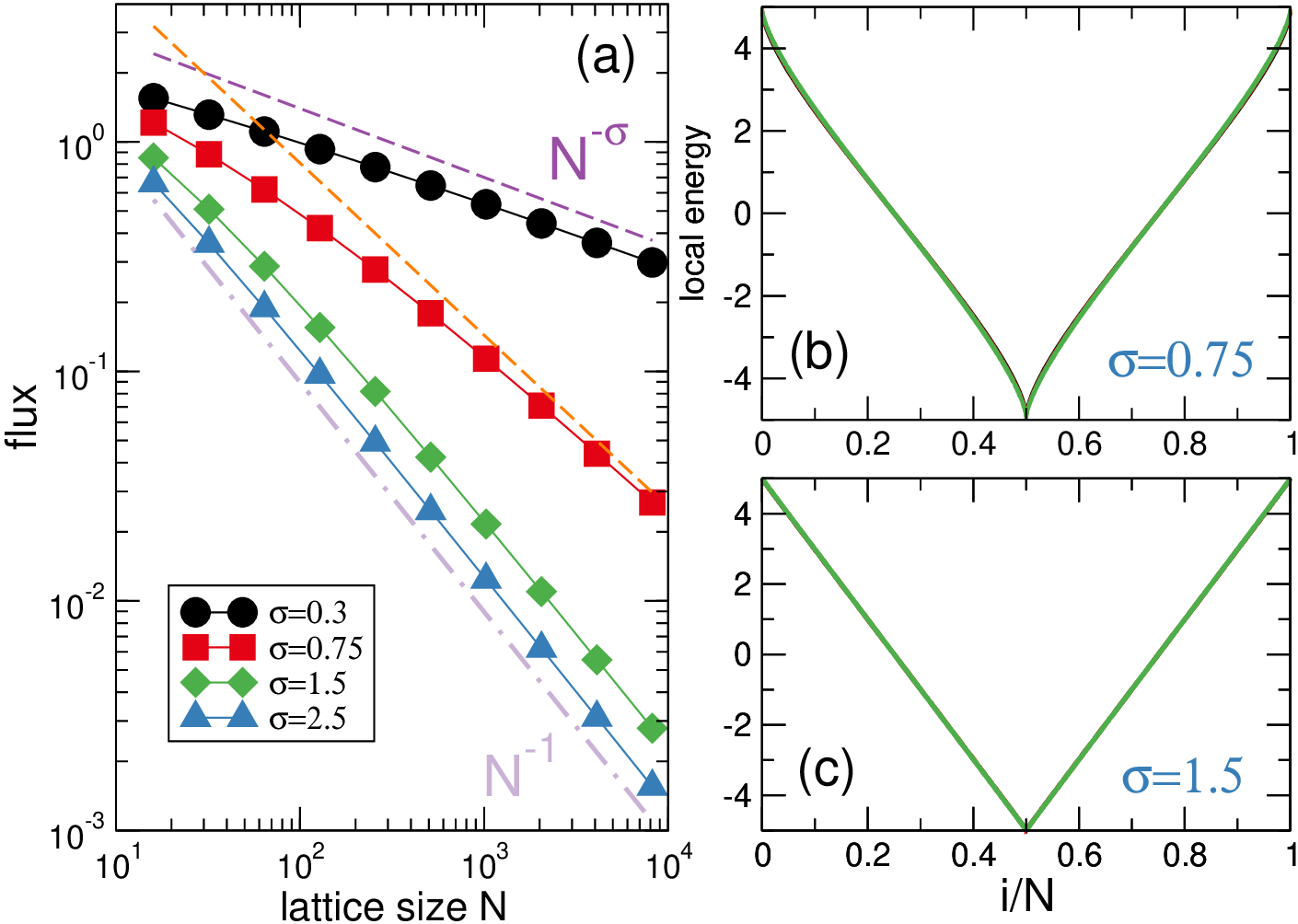}
\caption{Numerical solution of the L\'evy flight master equation
in non-equilibrium setup.  Equation (\ref{eq:ME}) has been solved in the steady state 
with a source and a sink located at sites $j=1,N/2$ respectively
and $\gamma=1,\epsilon=5$.
(a) Scaling of the steady state flux for different exponents showing
the anomalous and normal behavior for $\sigma<1$ and $\sigma>1$, respectively.
(b,c) Local energy profiles $h_i$ versus $i/N$  for two values of $\sigma$. Data for three lattice 
sizes $N=4096,8192,16384$ are plotted to demonstrate that large-size limit
is attained.}
\label{fig:levy}
\end{figure}

The results are given in fig.\ref{fig:levy}. To ease the comparison 
with the case of the $\varphi^4$ chain, we set $\nu=\sigma+2$ and report the 
results for a few values of $\sigma$, upon changing $N$. 
For large $N$ the flux scales as 
%\[
%Q\propto L^{-\sigma} \rm{for} 0<\sigma<1,  \quad
%Q\propto L^{-1} \rm{for } \sigma > 1
% \]
\begin{equation}
Q \;\propto\;  N^{-\sigma} \qquad \rm{for} \quad 0<\sigma<1, 
%Q & \propto & N^{-1}\quad \qquad \rm{for }\qquad \sigma > 1
\label{eq:Lsigma}
\end{equation} 
corresponding to superdiffusive transport, while normal transport
$Q \propto  N^{-1}$ occurs for $\sigma > 1$.
The case of a dynamical heat channel with L\'evy flight 
is discussed in \cite{Denisov03} but the prediction above is not given 
explicitly there.
  
It has been argued that standard linear response to a weak external force $F$ 
breaks down for  L\'evy flight dynamics~\cite{arkhincheev2001nonlinear}.
Indeed, the particle mobility is proportional to $F^{\nu-2}$ in the anomalous
regime. Equation (\ref{eq:Lsigma}) can be interpreted in the same way, considering the
applied thermal gradient $2\varepsilon/N$ as the (thermodynamic) force. 
Here, this result emerges in a many-body system due to the long-range interactions.
A similar conclusion has been drawn in \cite{Schukert2020} where the spin 
current of a spin chain has been shown to have the same scaling with respect 
to an external field.
Finally, the scaling (\ref{eq:Lsigma}) is also consistent with the fractional 
extension of the heat equation as discussed e.g. in \cite{li2020anomalous}.

\section{Nonequilibrium simulations}
It is now useful to directly inspect the nonequilibrium behaviour of the long-range $\varphi^4$ model~(\ref{eq:H}).
We consider an analogous transport setup where a $\varphi^4$ chain with periodic boundary conditions
interacts with two Maxwellian heat baths  placed at antipodal positions\footnote{As discussed in Sec.~\ref{sec:eq.},
the choice of periodic boundary conditions is motivated by the possibility to employ
 Fast Fourier Transform integration algorithms~\cite{Gupta2014} for the numerical integration of the equations of motion. 
 In the presence of long-range interactions and for large system sizes, these techniques can significantly increase the computational efficiency  with respect to standard routines.}. 
 Such reservoirs introduce elastic collisions with a model of ideal gas in equilibrium and impose two different temperatures
  $T_A\neq T_B$ in 	regions $A$ and $B$ of the chain.
 Collisions occur at random times  drawn independently from a Poissonian distribution $P(t)=\gamma_M \exp(-\gamma_M t)$, where $\gamma_M$ defines the strength of the coupling~\cite{LLP03,Iubini2018}. In between two consecutive collision events, the chain is evolved
 microcanonically with a symplectic fourth-order integration algorithm~\cite{mclachlan1992accuracy}.
In order to reduce boundary effects, we let each reservoir interact   with an extensive number $n_r=N/16$ of $\varphi^4$ lattice sites.
Accordingly,  at each collision time, $n_r$ independent collisions are generated on each thermalized region. 
Heat fluxes are measured as the time average of the total kinetic energy transferred from the reservoir to the particles belonging to the same thermalized region. More precisely, for region $A$ we define the flux
\begin{equation}
 Q_A=\overline{\sum_{i\in A} \delta K_i(t)}\,,
\end{equation}
where $\delta K_i(t)$ is the local variation of kinetic energy on site $i$ after a collision event and the bar denotes a time average. An analogous expression holds
for $Q_B$ and a nonequilibrium steady state is reached when $Q_A=-Q_B\equiv Q$. 

\begin{figure}
\hfill
\includegraphics[width=0.95\textwidth]{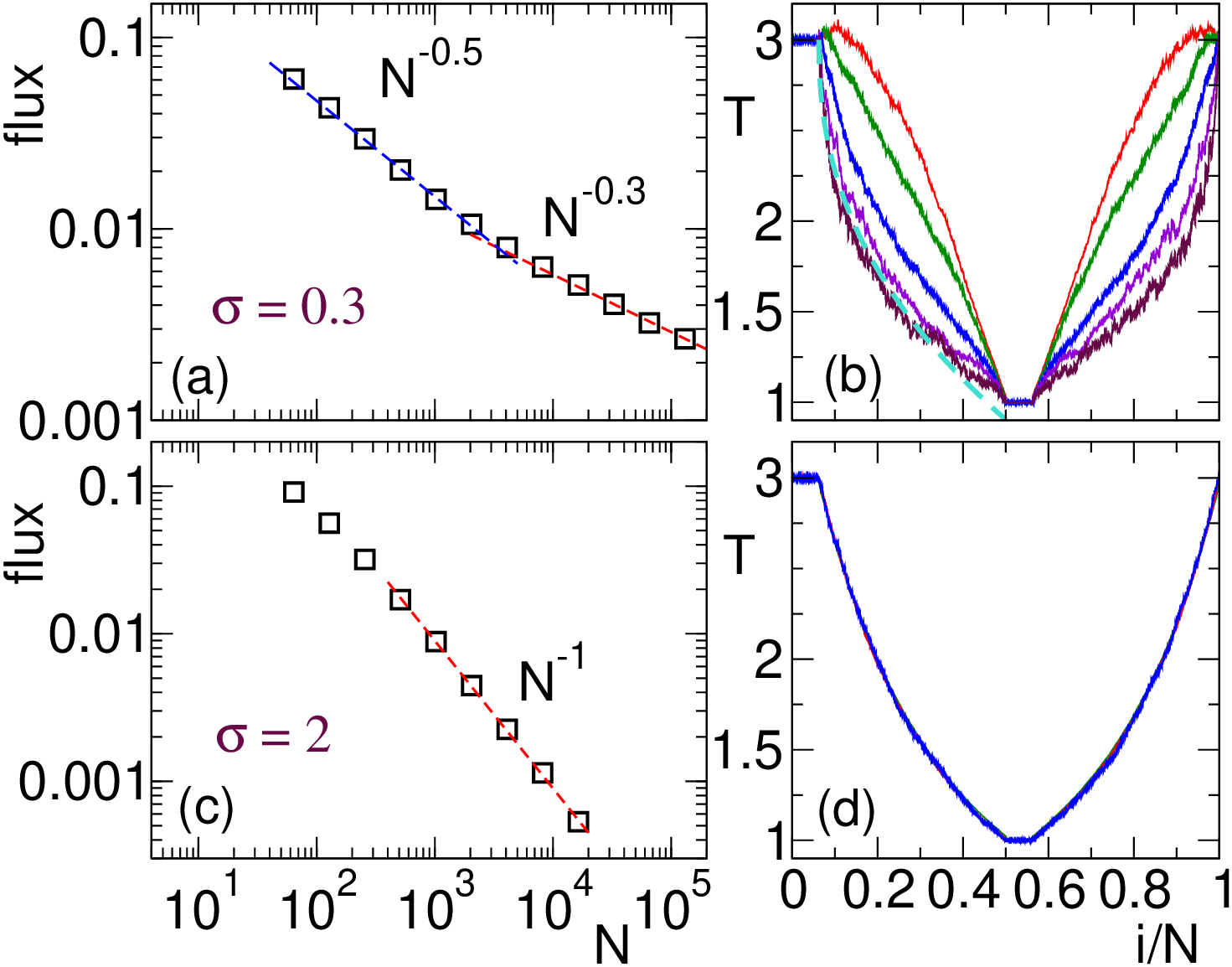}
\caption{
(a) Scaling of the stationary exchanged heat flux $Q$  versus lattice size for a $\varphi^4$ chain with $\sigma=0.3$ in contact with two reservoirs with $T_A=3$ and $T_B=1$.
(b) The corresponding temperature profiles versus $i/N$: curves from top to bottom correspond to $N=1024$, 4096, 16384, 65536, 131072. The cyan dashed line is a one-parameter fit of the function $T(\tilde{y})=T_A(1-\tilde{y}^\delta)$ (see text) with $\delta\simeq 0.43$ and fixed $T_A=3$. 
(c) Exchanged heat flux for $\sigma=2$ and same thermal boundary conditions. (d) Temperature profiles for $\sigma=2$ and $N=1024$, 4096, 16384. Nonequilibrium observables are averaged over a time up to $2\cdot 10^7$ units after a transient of at least $10^6$ units. }
\label{fig:nneq}
\end{figure}

In the following, we restrict ourselves to the case of the single-well local potential with parameter $\mu=2$ 
%{\bf [[questo è il prefattore che ho considerato davanti al termine long range]]}
 and we choose $\gamma_M=1$. We have verified that
this choice of parameters allows to optimize heat fluxes for temperatures $T_A=3$ and $T_B=1$ considered in our simulations.
In Fig.~\ref{fig:nneq}(a) we show the scaling of the stationary flux versus the lattice size $N$ for $\sigma=0.3$. Data appear to be 
initially well represented a power law $Q\simeq N^{-0.5}$ for a rather broad range of sizes, while they subsequently bend to the scaling $Q\simeq N^{-\sigma}$, in agreement with the Lévy flight model. The crossover between the two scaling laws is found to occur for $N\simeq 5000$.
 This evidence confirms that finite-size effects have an important role in the study of transport
properties of long-range systems, as discussed in~\cite{Iubini2018}.
 
 Further information is obtained from the shape of the stationary profiles of temperature measured along the chain 
 and reported as functions of the rescaled spatial variable $y=i/N$, see Fig.~\ref{fig:nneq}(b). 
 To the best of our numerical accuracy, we have some evidence that the  profiles tend to approach
 a highly nonlinear curve in the limit of large $N$, although some deviations are still present for the largest
 sizes here investigated. The temperature drop in the first half of the chain for the largest size $N=131072$ (see the continuous bottom line) is well fitted by the function $T(\tilde{y})=T_A(1-\tilde{y}^\delta)$, where $\tilde{y}=(y-n_r/N)$ is 
 the shifted spatial variable  restricted to 
% $\tilde{y}\geq 0$
  $n_r/N\leq y\leq 1/2$, see dashed line. The fitted value of the exponent
  $\delta \simeq 0.43$
 signals the presence of a singularity of the derivative of $T(\tilde{y})$ in $\tilde{y}=0$, i.e. at the boundary    with the thermalized region. On the other hand, finite-size effects deeply modify the shape of the profiles for smaller
sizes, see the upper solid line corresponding to $N=1024$, which exhibits a nearly vanishing first derivative close to $\tilde{y}=0$ and a negative second derivative.
      
Finally, in the region $\sigma>1$ numerical simulations confirm the occurrence of normal transport $Q\simeq N^{-1}$, as shown in Fig.~\ref{fig:nneq}(c) 
for $\sigma=2$.
The corresponding temperature profiles displayed in Fig.~\ref{fig:nneq}(d) are smooth and characterized by a very clean overlap, with a manifest reduction of finite-size effects with respect to the long-range regime.
Unlike the Lévy flight model (see Fig. \ref{fig:levy}(c)), they also manifest a clear deviation from the linear profile.
We argue that this is due to a slight dependence of the $\varphi^4$ diffusion constant on temperature in analogy with what observed in similar oscillators models~\cite{Iubini2016}.

\section{Discussion}

We have demonstrated that the fractional diffusion process and the 
L\'evy flight model well account
for the large-scale energy fluctuations of the long-range $\varphi^4$ chain. In hindsight, this may
seem trivial in view of the very fact that the couplings 
decay algebraically with the distance. However 
this is not the case for at least two reasons. 
First of all, the correct exponent is 
given by (\ref{es:sigma2}), a relation which was not foreseen
a priori. Second, models with the same   $r^{-1-\sigma}$ interactions
like the Hamiltonian XY model \cite{Olivares2016} or the Fermi-Pasta-Ulam-Tsingou model 
\cite{dicintio2019equilibrium} have 
different dynamical exponents.
In other words, models having the same coupling $r^{-1-\sigma}$
may belong to different dynamical universality classes, having
different hydrodynamics. 

%\item 
This conclusion is confirmed also by noting the differences between our model 
and the momentum-exchange model
with long-range interactions, recently studied in \cite{tamaki2020energy}.
The Authors considered the Hamiltonian (\ref{HLongRange}) with 
a quadratic pinning potential perturbed by a random exchange of momenta between the
nearest neighbor sites, occurring with a give rate.
The dependence of the exponent ruling the decay of energy current autocorrelation
(and thus the finite-size scaling of the heat flux in the nonequilibrum
setup) has a dependence on $\sigma$ which is different from ours.  
In their model superdiffusive transport occurs for $\frac12<\sigma<\frac32$
while here it happens for $0<\sigma<1$.
Also, even if their model leads to a fractional diffusion in the hydrodynamic limit \cite{suda2019family} which is the same as (\ref{eq:fdif}),	 the dependence 
of the order of the fractional derivative is different.
We thus argue that the two models belong to different classes.

%\item  
As far as scaling is concerned, 
anomalous energy diffusion appears not to be affected by the
fact that the $\varphi^4$ with a double-well potential undergoes the phase transition.  
As shown in fig. \ref{fig:dw}, even very close to the critical point the hydrodynamic behavior of the 
energy field still follows the fractional diffusion equation.
From our study we cannot exclude a sensitive dependence of thermal 
coefficients like e.g. $D_\sigma$ on energy around criticality,  
as seen for instance in the 
Ising model \cite{harris1988thermal,saito1999transport}.
Nonetheless, this observation is noteworthy and requires a theoretical
explanation. It has already been argued in \cite{staniscia2019differences} that for
Hamiltonian dynamics there is a coupling between the local
temperature field and order parameter fields.
So one might wonder how to include this effect in the hydrodynamics. 
The standard approach \cite{hohenberg1977theory} amounts 
to add to the dynamics of $m$ the one 
of the energy $h$, which is conserved and thus slow regardless
of the closeness to the critical point. 
This leads to Model C (see e.g. eqs. (4.50) in \cite{hohenberg1977theory})
that includes a conserved diffusive field. Its physical interpretation 
is that slow energy/temperature fluctuations 
modulate the coefficient $a_2$ in the Ginzburg-Landau functional.
It could be argued that the same would occur in the long-range case
and this would require adding to eq. (\ref{eq:fdif}) an 
equation for $m$ that accounts for the hydrodynamic 
behavior of the magnetization in the low-temperature phase.
This should be subject of future studies.

%\item 
Finally, in the present work we focused on equilibrium correlations
and on steady-state transport. One may thus wonder which 
is the connection with transient non-equilibrium problems. An important example in this context
is the phenomenon of coarsening dynamics, which has been studied 
in detail for the one-dimensional Ising model with long-range 
interactions (see \cite{corberi2019universality} and references
therein for a recent account) and for the present model 
\cite{staniscia2019differences}. Actually the dynamical 
exponent measured by equilibrium correlations 
is not related to the one 
characteristic of coarsening dynamics as measured in \cite{staniscia2019differences}.
The latter is essentially
determined by kinetics of domain walls \cite{corberi2019universality}. 

%\end{itemize}

\section{Summary and conclusions}

We have reported a simulation study of the discretized $\varphi^4$ lattice 
theory in one dimension, with long-range interactions decaying as an inverse power $r^{-1-\sigma}$ of the intersite distance $r$, and $\sigma>0$. 
We presented a dynamical scaling analysis of the relevant observable,  
energy structure factors and excess energy correlations.
Large-scale fluctuations of the local energy field display hydrodynamic 
behavior, which is diffusive for $\sigma>1$ and superdiffusive for $0<\sigma<1$
in both the cases with single and double-well
local potentials, with either attractive and repulsive couplings.
In the superdiffusive case, numerical data suggest that hydrodynamic
correlations are  
described by a fractional diffusion equation of 
order $\sigma$, eq. (\ref{eq:fdif}).
In the case of the double-well potential with attractive interaction
such behavior of energy fluctuations appear to be insensitive to the 
phase transition.
To further support the interpretation of the correlation data, 
we have successfully compared
both equilibrium and nonequilibrium simulations with an effective
model of transport based on L\'evy flights of energy carriers.
Altogether we conclude that the $\varphi^4$ 
model belongs to a different dynamical universality 
class with respect to others systems studied in the 
literature so far. We expect that the same conclusions apply to 
the class of Hamiltonians (\ref{eq:H}) with different 
local potentials $U(q)$. Our data will be of help to understand the 
large-scale properties of long-range interacting classical and 
quantum many-body systems.

\ack

SL acknowledges support from the program 
\textit{Collaborations of excellence in research and education} granted by SISSA 
(Trieste, Italy) where this work has been initiated. This work is part of the
MIUR-PRIN2017
project \textit{Coarse-grained description for non-equilibrium systems and transport phenomena} (CO-NEST)
No. 201798CZL.
We thank N. Defenu, G. Giachetti, P. Politi and A. Trombettoni for useful
discussions.

\section*{References}
\bibliography{slbib}

\providecommand{\newblock}{}
\begin{thebibliography}{10}
\expandafter\ifx\csname url\endcsname\relax
  \def\url#1{{\tt #1}}\fi
\expandafter\ifx\csname urlprefix\endcsname\relax\def\urlprefix{URL }\fi
\providecommand{\eprint}[2][]{\url{#2}}

\bibitem{LLP03}
Lepri S, Livi R and Politi A 2003 {\em Phys. Rep.\/} {\bf 377} 1

\bibitem{DHARREV}
Dhar A 2008 {\em Adv. Phys.\/} {\bf 57} 457--537

\bibitem{Lepri2016}
Lepri S (ed) 2016 {\em Thermal transport in low dimensions: from statistical
  physics to nanoscale heat transfer\/} ({\em Lect. Notes Phys\/} vol 921)
  (Springer-Verlag, Berlin Heidelberg)

\bibitem{Benenti2020}
Benenti G, Lepri S and Livi R 2020 {\em Frontiers in Physics\/} {\bf 8} 292
  \urlprefix\url{https://www.frontiersin.org/article/10.3389/fphy.2020.00292}

\bibitem{Zaburdaev2015}
Zaburdaev V, Denisov S and Klafter J 2015 {\em Reviews of Modern Physics\/}
  {\bf 87} 483

\bibitem{Cipriani05}
Cipriani P, Denisov S and Politi A 2005 {\em Phys. Rev. Lett.\/} {\bf 94}
  244301

\bibitem{Lepri2011}
Lepri S and Politi A 2011 {\em Phys. Rev. E\/} {\bf 83} 030107

\bibitem{Dhar2013}
Dhar A, Saito K and Derrida B 2013 {\em Phys. Rev. E\/} {\bf 87} 010103

\bibitem{Lepri2010}
Lepri S, Mej{\'{\i}}a-Monasterio C and Politi A 2010 {\em Journal of Physics A:
  Mathematical and Theoretical\/} {\bf 43} 065002
  \urlprefix\url{https://doi.org/10.1088/1751-8113/43/6/065002}

\bibitem{basile2016thermal}
Basile G, Bernardin C, Jara M, Komorowski T and Olla S 2016 Thermal
  conductivity in harmonic lattices with random collisions {\em Thermal
  transport in low dimensions\/} (Springer) pp 215--237

\bibitem{Cividini2017}
Cividini J, Kundu A, Miron A and Mukamel D 2017 {\em J. Stat. Mech: Theory
  Exp.\/} {\bf 2017} 013203

\bibitem{dhar2019anomalous}
Dhar A, Kundu A and Kundu A 2019 {\em Frontiers in Physics\/} {\bf 7} 159

\bibitem{Spohn2014}
Spohn H 2014 {\em J. Stat. Phys.\/} {\bf 154} 1191--1227

\bibitem{Mendl2013}
Mendl C~B and Spohn H 2013 {\em Phys. Rev. Lett.\/} {\bf 111}(23) 230601

\bibitem{Das2014}
Das S, Dhar A and Narayan O 2014 {\em J. Stat. Phys.;\/} {\bf 154} 204--213
  ISSN 0022-4715

\bibitem{mendl2015current}
Mendl C~B and Spohn H 2015 {\em Journal of Statistical Mechanics: Theory and
  Experiment\/} {\bf 2015} P03007

\bibitem{lepri2020too}
Lepri S, Livi R and Politi A 2020 {\em Physical Review Letters\/} {\bf 125}
  040604

\bibitem{Torcini1997}
Torcini A and Lepri S 1997 {\em Phys. Rev. E\/} {\bf 55} R3805

\bibitem{Metivier2014}
M{\'e}tivier D, Bachelard R and Kastner M 2014 {\em Phys. Rev. Lett.\/} {\bf
  112} 210601

\bibitem{avila2015length}
{\'A}vila R~R, Pereira E and Teixeira D~L 2015 {\em Physica A: Statistical
  Mechanics and its Applications\/} {\bf 423} 51--60

\bibitem{Olivares2016}
Olivares C and Anteneodo C 2016 {\em Phys. Rev. E\/} {\bf 94}(4) 042117

\bibitem{Bagchi2017}
Bagchi D 2017 {\em Phys. Rev. E\/} {\bf 95} 032102

\bibitem{bagchi2017energy}
Bagchi D 2017 {\em Phys. Rev. E\/} {\bf 96} 042121

\bibitem{Iubini2018}
Iubini S, Di~Cintio P, Lepri S, Livi R and Casetti L 2018 {\em Phys. Rev. E\/}
  {\bf 97}(3) 032102

\bibitem{wang2020thermal}
Wang J, Dmitriev S~V and Xiong D 2020 {\em Physical Review Research\/} {\bf 2}
  013179

\bibitem{Bouchet2010}
Bouchet F, Gupta S and Mukamel D 2010 {\em Physica A: Statistical Mechanics and
  its Applications\/} {\bf 389} 4389--4405

\bibitem{RuffoRev}
Campa A, Dauxois T and Ruffo S 2009 {\em Phys. Rep.\/} {\bf 480} 57--159

\bibitem{Campa2014}
Campa A, Dauxois T, Fanelli D and Ruffo S 2014 {\em Physics of long-range
  interacting systems\/} (OUP Oxford)

\bibitem{deBuyl2013}
de~Buyl P, De~Ninno G, Fanelli D, Nardini C, Patelli A, Piazza F and Yamaguchi
  Y~Y 2013 {\em Physical Review E\/} {\bf 87} 042110

\bibitem{hohenberg1977theory}
Hohenberg P~C and Halperin B~I 1977 {\em Reviews of Modern Physics\/} {\bf 49}
  435

\bibitem{harris1988thermal}
Harris R and Grant M 1988 {\em Physical Review B\/} {\bf 38} 9323

\bibitem{saito1999transport}
Saito K, Takesue S and Miyashita S 1999 {\em Physical Review E\/} {\bf 59} 2783

\bibitem{colangeli2018nonequilibrium}
Colangeli M, Giardina C, Giberti C and Vernia C 2018 {\em Physical Review E\/}
  {\bf 97} 030103

\bibitem{bermudez2013controlling}
Berm{\'u}dez A, Bruderer M and Plenio M~B 2013 {\em Physical review letters\/}
  {\bf 111} 040601

\bibitem{ramm2014energy}
Ramm M, Pruttivarasin T and H{\"a}ffner H 2014 {\em New Journal of Physics\/}
  {\bf 16} 063062

\bibitem{richerme2014non}
Richerme P, Gong Z~X, Lee A, Senko C, Smith J, Foss-Feig M, Michalakis S,
  Gorshkov A~V and Monroe C 2014 {\em Nature\/} {\bf 511} 198--201

\bibitem{moleron2019nonlinear}
Moler{\'o}n M, Chong C, Mart{\'\i}nez A~J, Porter M~A, Kevrekidis P~G and
  Daraio C 2019 {\em New Journal of Physics\/} {\bf 21} 063032

\bibitem{defenu2021long}
Defenu N, Donner T, Macr{\`\i} T, Pagano G, Ruffo S and Trombettoni A 2021 {\em
  arXiv preprint arXiv:2109.01063\/}

\bibitem{staniscia2019differences}
Staniscia F, Bachelard R, Dauxois T and De~Ninno G 2019 {\em EPL (Europhysics
  Letters)\/} {\bf 126} 17001

\bibitem{hu2000heat}
Hu B, Li B and Zhao H 2000 {\em Physical Review E\/} {\bf 61} 3828

\bibitem{Aoki00}
Aoki K and Kusnezov D 2000 {\em Phys. Lett. A\/} {\bf 265} 250

\bibitem{Piazza2009}
Piazza F and Lepri S 2009 {\em Phys. Rev. B\/} {\bf 79}(9) 094306
  \urlprefix\url{https://link.aps.org/doi/10.1103/PhysRevB.79.094306}

\bibitem{krumhansl1975dynamics}
Krumhansl J and Schrieffer J 1975 {\em Physical Review B\/} {\bf 11} 3535

\bibitem{dyson1969existence}
Dyson F~J 1969 {\em Communications in Mathematical Physics\/} {\bf 12} 91--107

\bibitem{mukamel2009notes}
Mukamel D 2009 {\em arXiv preprint arXiv:0905.1457\/}

\bibitem{desai1978statistical}
Desai R~C and Zwanzig R 1978 {\em Journal of Statistical Physics\/} {\bf 19}
  1--24

\bibitem{dauxois2003clustering}
Dauxois T, Lepri S and Ruffo S 2003 {\em Communications in Nonlinear Science
  and Numerical Simulation\/} {\bf 8} 375--387

\bibitem{aizenman1988discontinuity}
Aizenman M, Chayes J, Chayes L and Newman C 1988 {\em Journal of Statistical
  Physics\/} {\bf 50} 1--40

\bibitem{kerimov1993absence}
Kerimov A 1993 {\em Journal of statistical physics\/} {\bf 72} 571--620

\bibitem{Gupta2014}
Gupta S, Campa A and Ruffo S 2014 {\em Journal of Statistical Mechanics: Theory
  and Experiment\/} {\bf 2014} R08001

\bibitem{mclachlan1992accuracy}
McLachlan R~I and Atela P 1992 {\em Nonlinearity\/} {\bf 5} 541

\bibitem{caiani1998hamiltonian}
Caiani L, Casetti L and Pettini M 1998 {\em Journal of Physics A: Mathematical
  and General\/} {\bf 31} 3357

\bibitem{Zhao06}
Zhao H 2006 {\em Phys. Rev. Lett.\/} {\bf 96} 140602

\bibitem{Li2015}
Li Y, Liu S, Li N, H{\"a}nggi P and Li B 2015 {\em New J. Phys.\/} {\bf 17}
  043064

\bibitem{metzler2000random}
Metzler R and Klafter J 2000 {\em Physics reports\/} {\bf 339} 1--77

\bibitem{AnotransBook08}
Klages R, Radons G and Sokolov I~M (eds) 2008 {\em Anomalous Transport:
  Foundations and Applications\/} (Wiley-VCH Verlag, Weinheim)

\bibitem{Schukert2020}
Schuckert A, Lovas I and Knap M 2020 {\em Phys. Rev. B\/} {\bf 101}(2) 020416
  \urlprefix\url{https://link.aps.org/doi/10.1103/PhysRevB.101.020416}

\bibitem{Denisov03}
Denisov S, Klafter J and Urbakh M 2003 {\em Phys. Rev. Lett.\/} {\bf 91} 194301

\bibitem{arkhincheev2001nonlinear}
Arkhincheev V 2001 Nonlinear relation between diffusion and conductivity for
  {L\'evy} flights {\em AIP Conference Proceedings\/} vol 553 (American
  Institute of Physics) pp 231--235

\bibitem{li2020anomalous}
Li S~N and Cao B~Y 2020 {\em Applied Mathematics Letters\/} {\bf 99} 105992

\bibitem{Iubini2016}
Iubini S, Lepri S, Livi R and Politi A 2016 {\em New J. Phys.\/} {\bf 18}
  083023

\bibitem{dicintio2019equilibrium}
Di~Cintio P, Iubini S, Lepri S and Livi R 2019 {\em Journal of Physics A:
  Mathematical and Theoretical\/} {\bf 52} 274001

\bibitem{tamaki2020energy}
Tamaki S and Saito K 2020 {\em Physical Review E\/} {\bf 101} 042118

\bibitem{suda2019family}
Suda H 2019 {\em arXiv preprint arXiv:1912.01753\/}

\bibitem{corberi2019universality}
Corberi F, Lippiello E and Politi P 2019 {\em Journal of Statistical Mechanics:
  Theory and Experiment\/} {\bf 2019} 074002

\end{thebibliography}
\bibliographystyle{iopart-num}

\end{document}